\documentclass[10pt,journal]{IEEEtran}

\usepackage[nocompress,space]{cite}
\usepackage{csquotes}

\usepackage[notextcomp]{stix}
\usepackage{mathtools}
\usepackage{amsmath}
\usepackage{algorithm}
\usepackage[noend]{algpseudocode}
\usepackage{balance}
\usepackage[amsthm]{ntheorem}
\theoremstyle{definition}
\newtheorem{problem}{Problem}
\newtheorem{property}{Property}
\newtheorem{definition}{Definition}
\newtheorem*{hypothesis*}{Hypothesis}
\usepackage{listings}
\usepackage{url}
\usepackage{flushend}

\usepackage{xcolor}

\usepackage{graphicx}
\usepackage{tikz}
\usepackage{standalone}
\usetikzlibrary{calc,patterns,decorations.pathmorphing,decorations.markings}
\usetikzlibrary{positioning}
\usetikzlibrary{shapes,decorations}
\usetikzlibrary{shapes.geometric, arrows, positioning}
\usepackage{pgfplots}
\pgfplotsset{compat=1.14}
\usepackage[space]{cite}
\usepackage{xcolor}
\usepackage{todonotes}
\usepackage{booktabs}

\begin{document}

\title{Data Driven  Vulnerability Exploration\\
  for Design Phase System Analysis}

\author{Georgios~Bakirtzis, %~\IEEEmembership{Student Member,~IEEE,}
  Brandon~J.~Simon, %~\IEEEmembership{Student Member,~IEEE,}
  Aidan~G.~Collins, %~\IEEEmembership{Student Member,~IEEE,}
  Cody~H.~Fleming, %~\IEEEmembership{Member,~IEEE,}
  and~Carl~R.~Elks%,~\IEEEmembership{Member,~IEEE}% <-this % stops a space
  \IEEEcompsocitemizethanks{\IEEEcompsocthanksitem G. Bakirtzis and C.H. Fleming are with the University of Virginia, Charlottesville, VA USA. %
    E-mail: \{bakirtzis,fleming\}@virginia.edu
    \IEEEcompsocthanksitem B.J. Simon, A.G. Collins, and C.R. Elks are with Virginia Commonwealth University, Richmond, VA USA. %\protect\\
    Email: \{simonbj,collinsag,crelks\}@vcu.edu}% <-this % stops an unwanted space
  % \thanks{Manuscript received April 19, 2005; revised August 26, 2015.}%
}

\maketitle

  \begin{abstract}
    Applying security as a lifecycle practice is becoming increasingly important
    to combat targeted attacks in safety-critical systems.
    Among others there are two significant challenges in this area:
    (1) the need for models that can characterize a realistic system
    in the absence of an implementation and
    (2) an automated way to associate attack vector information; that is, historical data, to such system models.
    We propose the cybersecurity body of knowledge (CYBOK),
    which takes in sufficiently characteristic models
    of systems and acts as a search engine
    for potential attack vectors.
    CYBOK is fundamentally an algorithmic approach to vulnerability exploration, which is a significant extension to the body of knowledge it builds upon.
    By using CYBOK, security analysts and system designers can work together to assess the overall security posture of systems early in their lifecycle, during major design decisions and before final product designs. Consequently, assisting in applying security earlier and throughout the systems lifecycle.
  \end{abstract}

\section{Introduction}
\label{sec:org0b008b2}

It has been estimated that 70\%
of security flaws are introduced prior
to coding, most of which are due to the traditional practice
of application developers sharing
and reusing third party, legacy software---that
is assumed to be reasonably secure and trustworthy.
These flaws usually end up in the application software
and not, as might be expected, in network-based software \cite{fong2007web,third,safecode}.
Most security flaws are introduced as early design
or development decisions.

Both the academic and practicing cybersecurity community agree
that security engineering and analysis
as a full lifecycle practice, especially early
in the design process allows better awareness
and leverage at managing the challenges surrounding the unintentional introduction
of security flaws into complex systems.
This is especially important in the domain
of cyber-physical systems (CPS), where the exploitation
of software flaws and hardware weaknesses---introduced
by either importing software of unknown pedigree, incomplete security specifications,
or general unawareness of security characteristics
of given software, firmware, and/or hardware---can lead
to unforeseen physical behaviors that have consequences
in terms of safety, loss of vital service, and other societal impacts.

As modern CPS evolve into tightly integrated, extensible,
and networked entities, we significantly increase the attack surface
of these systems.
CPS now routinely employ a wide variety of networks, for example,
cloud, mobile services, industrial, internet of things
to realize a range of applications
from real time data analytics to autonomous vehicles control.
The use of extensible operating systems
and software to update code through loadable device drivers enhances productivity,
but it exposes the system to considerable risks
from attack injections.

With these insights and observations, we posit
that secure system design and deployment requires (1) planning
for cybersecurity from the outset as a strategic lifecycle activity,
and (2) taking the attackers perspectives
to best understand how to defend a system
from threats and exposing weaknesses before they become  vulnerabilities.
To achieve this goal methods and tools are needed
to allow security assessment throughout the systems lifecycle
and especially at the concept development phase,
where decision effectiveness is highest \cite{frola_system_1984,strafaci_what_2008}.

In recent years, a promising and rapidly growing approach
to enhancing awareness and managing challenges
of cybersecurity flaws
in evolving complex CPS is model-based engineering~\cite{nguyen:2017}.
Model-based analysis is firmly entrenched
in safety, dependability, and reliability engineering world
as evidenced  by such standards as IEC 61508
and ISO 26262, however model-based engineering is a late comer
to security \cite{nicol_model-based_2004}.

Models are generally treated as \emph{living documents}
maintained to reflect design choices and system revisions.
These models can be a valuable resource
for the security specialist
by providing what IT professionals
consider the ``what's''; that is, the rationale behind design choices
and not simply the resulting architecture of a system~\cite{chapman_what_2001}.

An additional benefit of model-based security is it tends
to look at security from a strategic point
of view, which means it attempts
to secure a system based on its expected service.
Rather than beginning with tactical questions
of how to protect a system against attacks, a strategic approach begins
with questions about what essential services
and functions must be secured against disruptions
and what represents unacceptable losses.
This is critical for CPS where losses
or disruptions to service can have dire societal
or safety impacts~\cite{alemzadeh:2013,kshetri:2017}.

\begin{figure*}[!t]
  \centering
  \includegraphics[width=.8\linewidth]{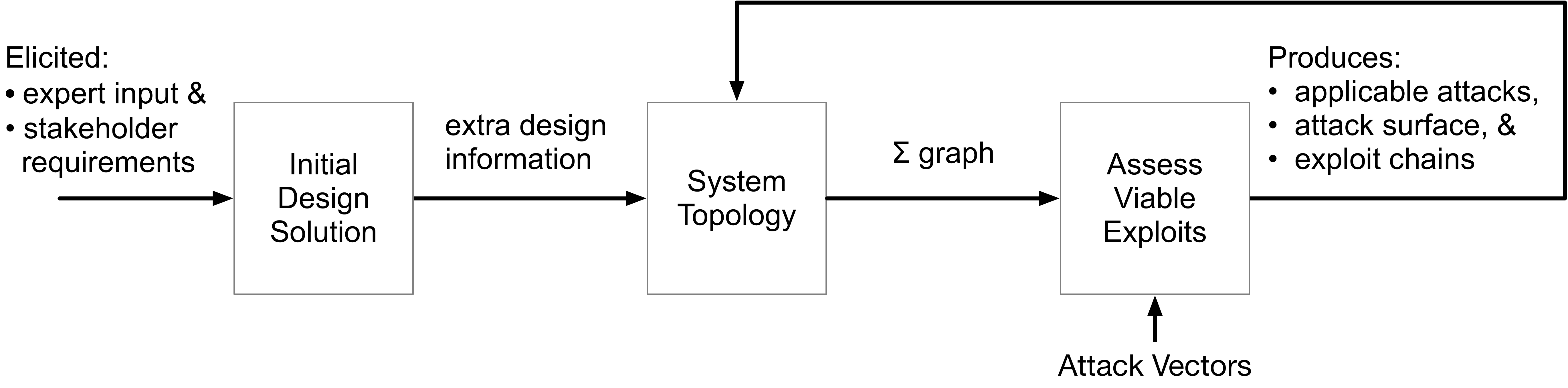}
  \caption{CYBOK depends upon stakeholder requirements and expert input
            to construct the initial design solution. Then, extra design information is added by a system designer. CYBOK takes as input
            the graph representation of that design and descriptions
            of attack vectors to produce all applicable attacks throughout the topology of the system, the attack surface of the system, and
            the exploit chains traversing the system topology.}
  \label{fig:process}
\end{figure*}

However, one of the major impediments
to effectively transition security assessment
into the model-based engineering realm has been
associating system models to applicable attack vectors.
These models reside in a higher-level of abstraction
than what is typically present in cybersecurity analysis.
Our aim is to use the model to drive the attack vector analysis
in the design phase.
There are two things necessary to achieve this congruence:
(1) understand the data available to security researchers
and decide on which of those can inform early
on and (2) capture lower-level information
in the model, such that it can be used
to associate the available data with the model.
Such an approach bridges the gap
between existing curated attack vector information
and models of systems.
Indeed, this paper presents one answer
to examining cybersecurity concerns
in the model-based engineering setting.

Towards this goal we have previously~\cite{bakirtzis2018model} presented a CPS model that includes a schema with extra design information to manually associate attack vector data describing attack patterns, weaknesses, and vulnerabilities. The previous work proposed just a model, not an algorithmic solution to the vulnerability exploration problem. To explore the large amount of data compiled by security professionals, it is helpful to associate attack vectors algorithmically. This is precisely the topic of this paper.

\textbf{Contributions.} The contributions of this work are:
\begin{itemize}
\item an algorithmic implementation, called cybersecurity body of knowledge (CYBOK),
  that accepts as input such models
  and produces:
  \begin{itemize}
  \item a component-wise attack vector analysis using attack vector data,
  \item a notion of attack surface, which only depends on the model, and
  \item all exploit chains applicable to subsystems of the system model and
  \end{itemize}
\item a demonstration of the method on an unmanned aerial system (UAS).
\end{itemize}

% introduction: systems engineering cybersecurity.
% data informatics, what data do I need to use to inform early on
% model-based

\section{Model-Based Security Analysis}
\label{sec:org103a848}

% add stuff about model-based security analysis as an intro to this section
% how we differ, what do we think it should look like?
% state explicitely that this early-design phase analysis.

Model-based security analysis is a relatively new field
that attempts---as the name implies---to understand system threats
through the use of models.
In this paradigm, models are used either
as an augmentation to other security strategies
during deployment or as evidence
to support design decisions early
in the systems lifecycle.
However, most current models are probabilistic
in nature and, therefore, require a ground truth.
These models also heavily depend
on the modelers expertise and experience.
Any such model captures exactly that expertise
such that it is communicated to other stakeholders.
To our knowledge, models used at the design phase
have not achieved fidelity
with security data collected and otherwise used
in already realized systems,
which would consist of an important addition
to defending against increasingly sophisticated threats.

But why is that? To understand the difficulty
of finding vulnerabilities in system models---instead
of a deployed product---it is important
to first define the difference
between bugs, vulnerabilities, and exploits.
Successful exploits take advantage
of flaws (either serious design flaws
or unexpected system behavior that is implementation specific).
These flaws in the system are called bugs.
However, not all bugs are vulnerabilities.
Only a subset of bugs that can lead
to exploitation are vulnerabilities.
This notion leads to the first problem
that is addressed in this paper.

\begin{problem}
  Vulnerabilities are explicitly found
  at the level of code or hardware.
  However, to address system security early
  in the design cycle, there need to be methods
  that can identify potential vulnerabilities before code development.
\end{problem}

% Here is is my inspiration, insight into solving this problem
% "what's and how's
% abstract to general to specifics
To bridge the gap between models
and realized solutions requires
constructing an initial design
of the system.
This design needs to include
both the \emph{what's}, the components
of the system, for example GPS,
and the \emph{how's}; that is a particular
hardware, firmware, and software solution
that implements some desired function.
Additionally, any such model needs
to include the interaction between components
as is defined by their communication
and data transfer.

One way to fulfill those requirements is
to model CPS as a graph
of assets but with added information
in the form of descriptive keywords.
This is a reasonable and appropriate model
as it pertains to security analysis.
Attackers typically think in terms of graphs, through a series
of increasing violations based on concepts
of connectivity, reachability, and dependence,
not in lists of assets as---most commonly---defenders do \cite{lambert_defenders_2015}.

In addition, this model must include extra information
in the form of keywords that augment the model,
resulting in a system model that captures the choices a designer is considering about hardware, firmware, and software.
These augmentations can be done without overly specific details about its final implementation.
This is a key feature that reflects how designs evolve in the construction of a system, where choices
of specific hardware and software are done early in the development cycle.
Furthermore, the ease of changing those keywords
to describe a functionally equivalent system allows
for modeling flexibility that is not available after code has been written
and designs are finalized.

Formally, an architectural model of a CPS can be captured in a graph, \(\Sigma \triangleq \left(\mathcal{V}, \mathcal{E}, \mathcal{D}\right)\), where,

\[
  \mathcal{V} \triangleq \left\{v \mid v = \text{a system asset} \right\}\text{,}
\]
\[
  \mathcal{E} \triangleq \left\{e \mid e = (v_i, v_j); v_i, v_j \in \mathcal{V} \text{ dependent assets} \right\}\text{, and}
\]
\[
  \mathcal{D} \triangleq \left\{ d \mid d = \left(w_1, w_2, \dots, w_n\right) \text{ descriptive keywords}\right\}\text{.}
\]

% what is this model? a paragraph no more.

In order to extract and use the descriptive information; that is, the extra keywords,
from a given vertex or edge, we define the descriptor function,
$\text{\textit{desc}}: \mathcal{V} \cup \mathcal{E} \rightarrow \mathcal{D}$.

\begin{figure}[!t]
  \includegraphics[width=0.45\textwidth]{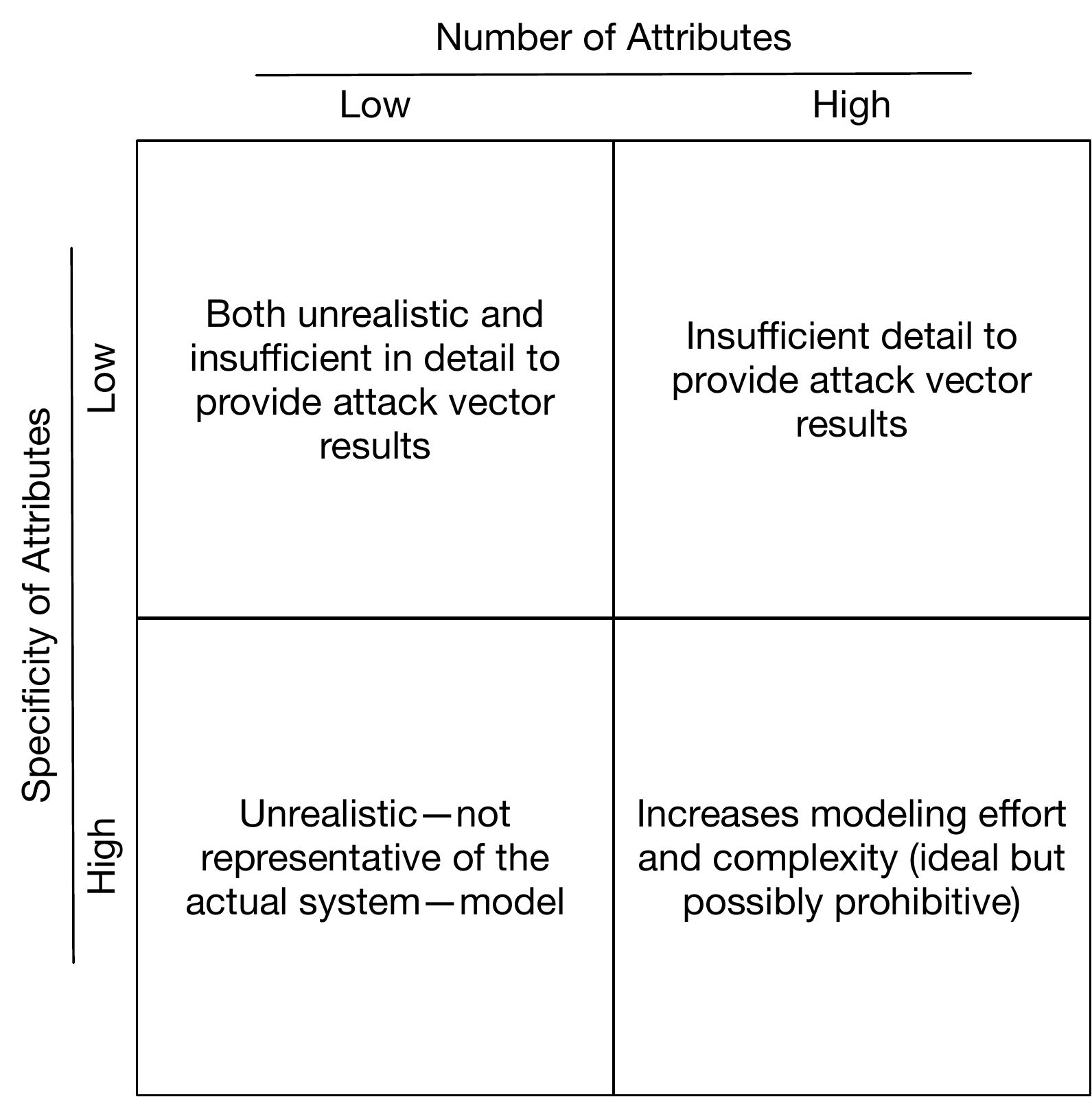}
  \caption{The fidelity of the attributes describing a CPS has to associate to the attack vector information (reproduced from Bakirtzis et al. \cite{bakirtzis2018model}).}
  \label{fig:grid}
\end{figure}

The above definitions lead to the first practical challenge, which is to determine if a model is sufficient
for security assessment (Fig.~\ref{fig:grid}).
It is important at these early design stages for a system model
to sufficiently describe functionally complete system---by adding hardware
and software information that,
if put together, implements the desired functional behaviors
expected from the system.

% clear challenges paragraph here or below
% these are going to be for any model so start maybe Any model in security analysis...

Any model used for security analysis contains two main challenges.
The first challenge, is the amount of data associated with the model.
When using a realized system it is possible
to mine all possible configuration settings including software
and firmware versions.
In the absence of the implementation such information can still be reflected in a modeling setting
but it requires significant modeling cost in terms of time and expertise.
It can also be less informative at the design phase than a more general model
because a slight change in versioning will hide a class
of weaknesses and vulnerabilities.
The second challenge, is the level of abstraction the model resides in.
No model is a direct reflection of a realized system
but any model needs to be specific enough
to be informative.
This is a difficult task and largely depends
on the given abstraction set overall by the modeling process
as well as the expertise of the modeller.

These are precisely the challenges that the added keywords address.
By changing the specificity and amount of keywords, we change the overall fidelity
of the system model, $\Sigma$.
It is through that \emph{extra} design information
that our solution, CYBOK, is able to take the graph of a system model
and map applicable attacks
from security databases (Fig~\ref{fig:process}).
While there may be a number of different criteria
for selection, in previous work we have found
that the extra design information can be categorized
through the following practical schema: operating
system, device name, communication, hardware, firmware,
software, and entry points~\cite{bakirtzis2018model}.
Each of the categories is expected to contain a string of keywords, \(d \in \mathcal{D}\)
that collectively describe a given system solution.
A given category can also duplicate the descriptive keywords
present in another category or simply contain the null set, \(\emptyset\).

The fidelity of the model is still based on the choices of the modeller.
On the one side of the \emph{model sufficiency spectrum} there are designs
that are too general and do not contain information about the system that would aid in determining security posture.
On the other side of the spectrum there are designs that are too specific,
to the extent that the effort to create
and potentially modify is equivalent
to constructing an actual implementation of the system and its functionality.
Such systems are complete but impractical.
There is a spot in the middle
of the spectrum, where the information contained
in the model can provide a reasonable idea about the system's threat space
without being so detailed it is inflexible and costly
to construct and maintain throughout its lifecycle.

A perhaps less obvious but equally important challenge refers
to the information necessary to associate the model, \(\Sigma\),
to potential attack vectors.

% this is an association problem with CVE, CWE, CAPEC...
% not structured and incomplete but how to associate it with the formal model...
\begin{problem}
  How can we associate vulnerability, weakness, and attack pattern information
  that is intended to be used by security analysts to a model?
  % How can attack vector information be used
  % that assumes it is deciphered, collected, and filtered by human analysts
  % and, therefore, is unstructured, incomplete, and sometimes ambiguous
  % with respect to an automated vulnerability analysis tool?
\end{problem}

This problem is difficult because
of the way security experts record vulnerabilities.
While several attempts have been made to standardize the form
of an attack vector entry, the current situation is such
that the different databases are based on a different schema,
because they rely
on the deduction and inference capabilities
of a human.
This means there is no straightforward approach
to feeding that data into a machine
to automatically find those mapping.

Our solution is based on the set of descriptors, \(\mathcal{D}\), present
in the model and using standard practices
from natural language processing
to deconstruct and associate the contents of an attack vector entry---in the form
of text---to the models keywords, \(d \in \mathcal{D}\).
This requires a separate set of attack vector entries, $\mathcal{AV} \triangleq \left\{ av \mid av = \text{a set of stemmed words}\right\}$. %= \left(k_1, k_2, \dots, k_n \right)\text{ descriptive keywords}\right\}$.
Therefore, the fundamental problem that CYBOK attempts to solve is then formalized as the function,
$\text{\textit{associate}}: \text{\textit{desc}} \rightarrow \mathcal{AV}$.
% We complement the analysis by using two additional metrics: the attack surface, \(\mathcal{AS}\),
% and exploit chains.
% Both of which follow from the information produces by the $\text{match}$ function.

By doing so, the problem is reduced
to associating keywords describing the system (which exist within the model)
to stemmed words describing attack vector information
(which are constructed using the contents
of the database entries and natural language processing).

Applying security consideration to model-based engineering is difficult for several reasons.
Three of the most important difficulties are:
the curation of information (both from the model
and the attack vector databases),
the intuition surrounding the fidelity of the system model,
and the development of an algorithmic approach
that allows for filtering through a large number
of attack vectors produced at the design phase.

Finding attack vectors for individual vertices
or edges overcomes these challenges.
However, it can be daunting to see the big picture
when confronted with such a larger amount of data.
Security professionals frequently use complimentary metrics
to understand the overall security posture
of a given system.

Two such useful metrics for security analysis are:
\begin{enumerate}
\item The \emph{attack surface} captures all the entry points
  into a given system \cite{manadhata_attack_2011}.
\item The potential for further spread; that is, further violations,
  after an element of the attack surface has been compromised,
  known as an \emph{exploit chain}.
\end{enumerate}
\vspace{1em}

% This means that while CYBOK can find all instances of attack vectors that match
% both vertices and edges,
% we only care about the entry points
% and subsequent violations from that point thereafter.
% This is to say that we consider only realistic attacks,
% within the system structure.
% A single system element, $v \in \mathcal{V}(\Sigma)$, that maps to an attack, $av \in AV$,
% based on some descriptive keyword, $d \in \mathcal{D}$, but not the entry point
% and does not connect to other subsystems that could potentially be violated
% at the entry point is
% considerably less likely to be violated than other, more accessible,
% vertices.
% Denoting the set of such vertices with \(\mathcal{S} \subset \mathcal{V}\) allows us
% to formalize the desired set, \(\mathcal{R} \triangleq \mathcal{V} \setminus \mathcal{S}\).

% \begin{hypothesis*}
%   The proposed solution, CYBOK,
%   will be able to find the elements of the attack surface
%   and calculate the spread from that point
%   to other system assets using evidence; that is, documented
%   and recorded attacks through models of systems.
%   In this case, the role
%   of CYBOK is to automatically choose attacks
%   from an existing set and see if they apply
%   to the model.
%   Hence, CYBOK simulates attacker behavior,
%   since a significant number of attackers
%   reuse attack patterns for separate objectives~\cite{allodi_work_2017}.
% \end{hypothesis*}

\noindent
\textbf{Proposed Solution}. \quad In general, CYBOK is an algorithmic solution
that takes as input a sufficient system model (of \(\Sigma\) form) to associate to the body of knowledge of attack vectors (of \(\mathcal{AV}\) form). By knowing the associated attack vectors it then produces security metrics only based on the model; that is, the attack surface of the system model and the exploit chains for a particular element of the model.\\

\noindent
\textbf{Intrinsic Limitations}. \quad Model-based security analysis is grounded
on early design information. This early design information is usually
incomplete and abstracted with respect to the final design solution.
This leads to result spaces; that is, associated attack vectors,
that are significantly larger than when analyzing a realized system.
Navigating through the results can be challenging for systems engineers
that are not familiar with security practices.
Our aim is to provide a framework in which security analysts and
systems engineers work synergistically to understand
both necessary design decisions (that might affect potential security mitigations)
and security considerations (that might affect the design of the system).

An additional limitation is the fidelity of the model.
The model can only be as good as the person who is modeling the system.
Therefore, a poorly constructed model might mislead instead
of providing insight into the security posture of the system.

\section{System Model}
\label{sec:org5ceca0b}
% \label{org470788e}

To address the challenges in the previous section and construct a sufficient model with respect to vulnerability analysis, first we must elicit information from the stakeholders. The stakeholders of the system include the owners of the eventual system, the system designers, the safety engineers, and the security analysts. While it is outside the scope of this paper to address the systematic process in which such information is elicited (see Carter et al.~\cite{carter:2018} for further details on the topic), it is important to place CYBOK within its larger framework. Without this framework it would not be possible to have complete or correct information to apply vulnerability exploration this early in the system's lifecycle. Based on this elicitation, an initial design solution is modeled in the systems modeling language (SysML).

% We established the usefulness
% and practicality of SysML as a specification language.
SysML uses visual representations
to capture the system design process through objects.
The main benefit of SysML is
that it presents the same information in different views,
which allows the same system to be modeled based
on its requirements (through the requirements diagram),
through its behavior (through, for example, the activity
and/or state machine diagrams), and/or through its architecture
(through block definition diagram (BDD) and/or internal block diagram (IBD)).

% kkl: metamodel/schema be explicit
To model CPS architectures
in SysML the system structure
is captured as a set of BDD and IBD.
The BDD view of the system shows the composition of the system.
The IBD view refines those compositions
to interconnections within the system
and how those interconnections compose the system behavior.

However, each element of the system model is described
by a standardized schema
as presented by Bakirtzis et al.~\cite{bakirtzis2018model}.
It is, therefore, not necessary to capture
this model in SysML.
This model is flexible
to design changes
and has supported vulnerability analysis
in a manual setting.
In this work we use the modeling methodology
to support automated vulnerability analysis.

Security specialists usually construct the following information
implicitly through expertise. To automate this task
this implicit information needs to be captured explicitly in the model.
This information will also assist in constructing a living document
describing the \textit{what's} of those choices.
To recap, the schema is composed
by the following categories that describe each system element:
\begin{itemize}
\item operating system,
\item device name,
\item communication,
\item hardware,
\item firmware,
\item software, and
\item entry points.
\end{itemize}

% In turn, each of the categories contains a string of keywords, \(d \in \mathcal{D}\).
% A given category can also duplicate the descriptive keywords
% present in another category or simply contain the null set, \(\emptyset\).
% This information is captured in SysML within the IBD as part properties.

It is through that \emph{extra} design information---gathered
by eliciting stakeholder information
and inspecting design documentation---that
CYBOK is able to take the graph of a system model
and map potential attacks
from databases (Section~\ref{sec:org27e0311}).
This is done by using the key terms presented in the schema
for each element and checking if they are present
in the documents composing the databases.

Reasoning in terms of the diagrams has several benefits
during the design process that hold
for security analysis in general (see Oates et al. \cite{oates_security_2013}), which is the benefit of starting with a SysML model instead of its graph representation.
This is less true for matching attack vectors
to the model.
The exporting of models in a standardized format is, therefore, beneficial.
The translation of IBD diagrams into a graph
is encoded into GraphML---a simple XML format
that is widely used to import and export graph structures~\cite{brandes2013graph}.

The two models---i.e., the visual representation in SysML and the graph structure----must be isomorphic. This means that the transformation between the SysML model and the GraphML representation must not change the model of the system. To achieve an isomorphic transformation we apply a model transformation on the IBD model. This transformation produces a sufficient graph model for security analysis (Fig.~\ref{fig:topology} in Section~\ref{sec:org7815955}).

\begin{property}[Model Transformation]
  An \textsc{internal block diagram} is the graph \(I \triangleq \left(\mathcal{V}, \mathcal{P}, \mathcal{D}\right)\),
  where \(\mathcal{V}\) is the set of vertices of \(I\) and \(\mathcal{P}\) is the set of ports of \(I\).
  Further, \(\mathcal{V}\) represents the assets of a cyber-physical system, \(\mathcal{P}\) the dependence between assets,
  and \(D\) the descriptors of each asset.
  Therefore, the graph \(I\) is isomorphic to our system graph \(\Sigma\); that is to say \(I \cong \Sigma\).
\end{property}

This system model represented as a graph allows us to think similarly to attackers
and to construct connections that would not be obvious
if treated as simple decoupled components.
The graph of the system model assists with not only finding
vulnerable subsystems individually but also with
finding the attack surface of the system and composing exploit chains.
Exploit chains are a subset of attacks that could traverse
though the system model graph and elevate the impact to deteriorate system behavior.
% This analysis contains some nuances. Namely, that a telling
% but incomplete security metric is the system's attack surface;
% that is, the exploitable entry points into the system structure.
All exploit chains start at elements of the attack surface.
% These are handled by the taxonomy by adding
% the ontological notion of design elements that are at the entry point
% of a subsystem.
The attack surface, \(\mathcal{AS}\) is composed
by all vertices that an attack from the databases is found
to be potentially applicable at the entry point description.

% GB: Develop these as they apply and how they apply over Σ.
In particular the graph model allows us
to define the attack surface and exploit chains
over the system model \(\Sigma\)
in a straightforward manner.

\begin{definition}[Attack Surface]
  We define the attack surface of a system model, $\Sigma$ 
  as \(\mathcal{AS} \subseteq \mathcal{V}\),
  which is composed by all vertices that can be entry points
  into the system and allow the attacker to cause further spread within the system structure.
\end{definition}

\begin{definition}[Exploit Chain]
  To construct the exploit chain we define a function,
  $\textit{paths}: \mathcal{AS} \times t \times \Sigma \rightarrow \mathcal{P}$,
  where $\mathcal{AS}$ the sources
  of all paths and \(t\) a used specified target and $\mathcal{P}$, the set
  of all simple paths from the source to the target over $\Sigma$.
  Then to construct a single exploit chain, $ec \in \mathcal{EC}$, the set \(\mathcal{P}\) is filtered
  by checking if every vertex and edge within each individual path associates to some attack vector
  from \(\mathcal{AV}\).
\end{definition}

\section{Attack Vector Dataset}
\label{sec:org27e0311}
% figure here, why do people use it, why is it the state of the practice

% Ideally, a set of databases that define and score the impact
% using several metrics --
% for example, resilience, safety of the system --
% would be more useful in this context.
% However, since these databases do not exist,
% we need to extrapolate and measure the posture
% of the system based on the curated information
% that does exist.
% Namely, this information comes in the form
% of attack vector databases encapsulating general attack patterns,
% weaknesses, and recorded, specific system vulnerabilities.
% In the model-based system engineering realm
% these databases can assist in understanding
% and refactoring design solutions that are more secure
% without increase in cost or complexity.

CYBOK is composed by several databases
to address two main challenges.
The first challenge is finding applicable attack vectors based on a system model.
The second challenge is to present a reasonable amount of data
to the security analyst, such that they can erect barriers
or add resilience solutions to strengthen the design of CPS
using an evidence-based approach.

\begin{figure}
  \centering
  \includegraphics[width=1\linewidth]{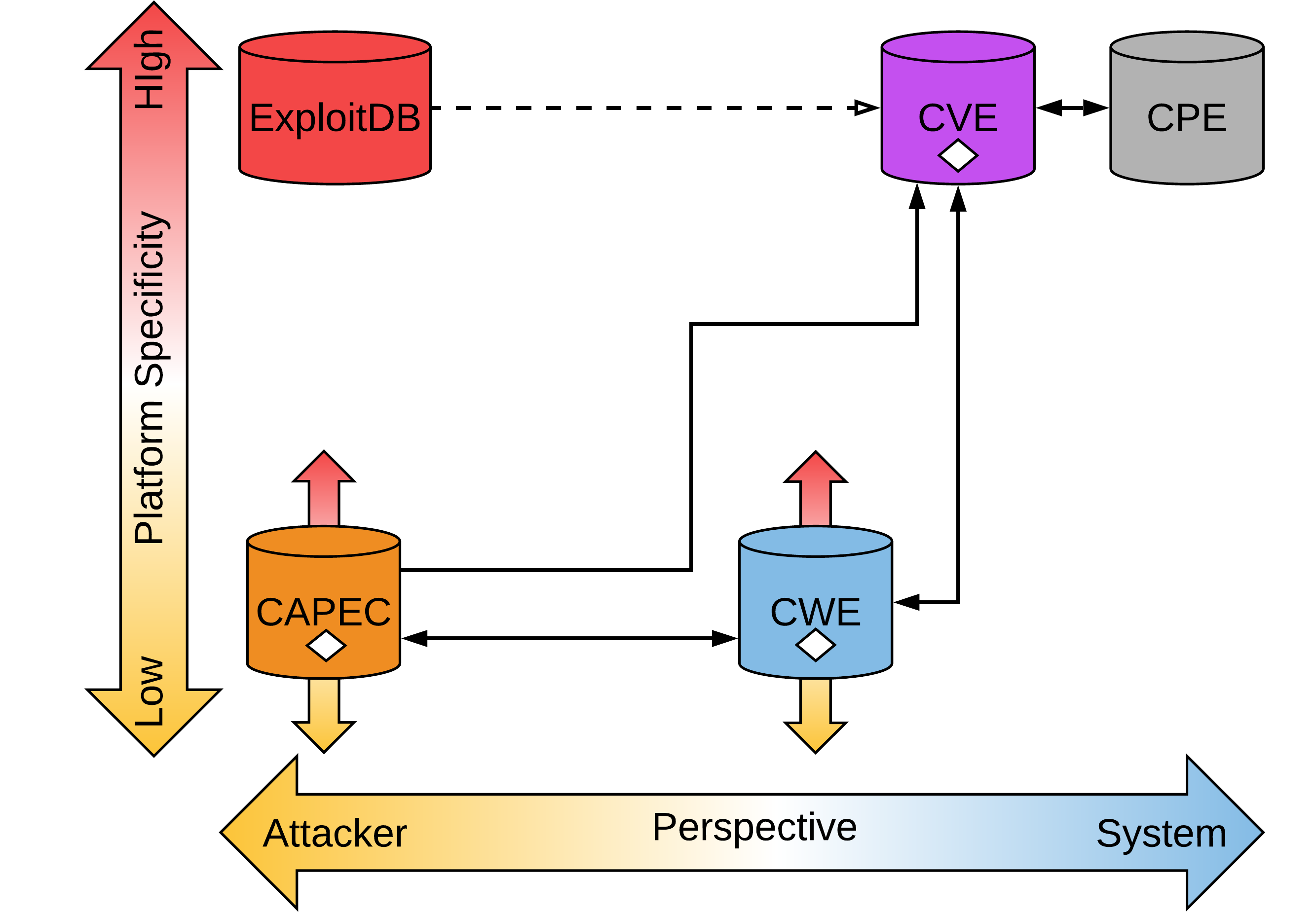}
  \caption{The collection of the most common open attack vector datasets and their interconnections in the sense of topical relationships with regard to attacker and defender perspectives and level of specificity they contain about platform information. Edges denote which datasets have explicit relationships with one another. CYBOK only uses a subset of these databases marked by a $\Diamond$. }
  \label{fig:datasets}
\end{figure}

To address these challenges, CYBOK incorporates three collections curated
by the MITRE corporation: CVE~\cite{CVE,NVD}, CWE~\cite{CWE}, and CAPEC~\cite{CAPEC}.
CVE is the lowest level of attack vector expression,
defining tested and recorded vulnerabilities on specific systems.
CWE presents a hierarchy of known system weaknesses at different levels of abstraction, from which exploits can be derived.
Finally, CAPEC provides a high level view of attacks against systems at varying levels of abstraction, in the form of a hierarchy organized by the goal or mechanism of each attack.
These three collections also include relationships
to one another (Fig.~\ref{fig:datasets}).
Formally CAPEC, CWE, and CVE construct the set
of attack patterns, weaknesses,
and vulnerabilities \(A \times W \times V \cong \mathcal{AV} \).

Other known datasets include the MITRE Common Platform Enumeration (CPE)~\cite{CPE} and the Exploit Database (exploit-db)~\cite{exploit-db}.
The former includes information that is platform specific, including particular versions of software that a particular exploit is associated with. The latter provides samples of exploits for given CVE entries. The reasons for not including these two datasets is because CPE requires knowing the specific versions of software that are going to be on the system---which are not known at design phase---and exploit-db requires a realized system to test exploits against.

Using CVE entries has the benefit
of finding additional attack vectors
because the specificity of their descriptions may more closely associate to the descriptions in the model than those of CAPEC and CWE.
When CVEs are matched to the system model, CYBOK uses that information to abstract upwards
towards the weaknesses and the attack patterns.
This is especially useful in the case where multiple CVE entries
are associated with a subsystem that has the same associated weakness
or attack pattern.

% Furthermore,
In general the CWE and CAPEC abstractions are more useful
to designers over CVE entries
because CVEs are too specific
to be useful at the design phase.
For example, being aware of a vulnerability in a specific version
of software is less illuminating than knowing that a specific class
of software bugs might consist of a vulnerability in the implementation
of the system---and therefore can construct more concrete requirements
or define specific mitigations.
On the other hand, a number of applicable attack vectors from the model are going
to reside in CVE.
At the same time, CVE contains a significantly larger number of entries than both CAPEC and CWE (\({\sim}100,000\) vs. \({\sim}1500\)), meaning the addition of CVE entries will explode the number
of results for a given system \(\Sigma\).
Therefore, all three are needed: CVE entries
to be more thorough and complete in analyzing the system model and CAPEC, CWE
to abstract to useful information to system designers.

% \begin{figure}[!t]
%   \centering
%   \includegraphics[width=\linewidth]{./figures/db.tex}
%   \caption{CVE is two magnitudes larger than both CWE and CAPEC.}
%   \label{fig:db}
% \end{figure}

% Other databases could be included as long as they are congruent
% to the same ontology.

While the three selected collections
are complete in relation to the sufficient model,
CVE is certainly not exhaustive.
There are certainly instances where companies
or government agencies curate more exhaustive databases
from their non-disclosed findings.
In those cases, the design of CYBOK can be extended
to use the information contained
in these private databases.

\section{Architecture}
\label{sec:org51a89ab}
% necessary functionality to solve figure 2.
% what are all the thing needed to do the job of the first three pages.

\begin{figure}[!t]
  \centering
  \includegraphics[width=\columnwidth]{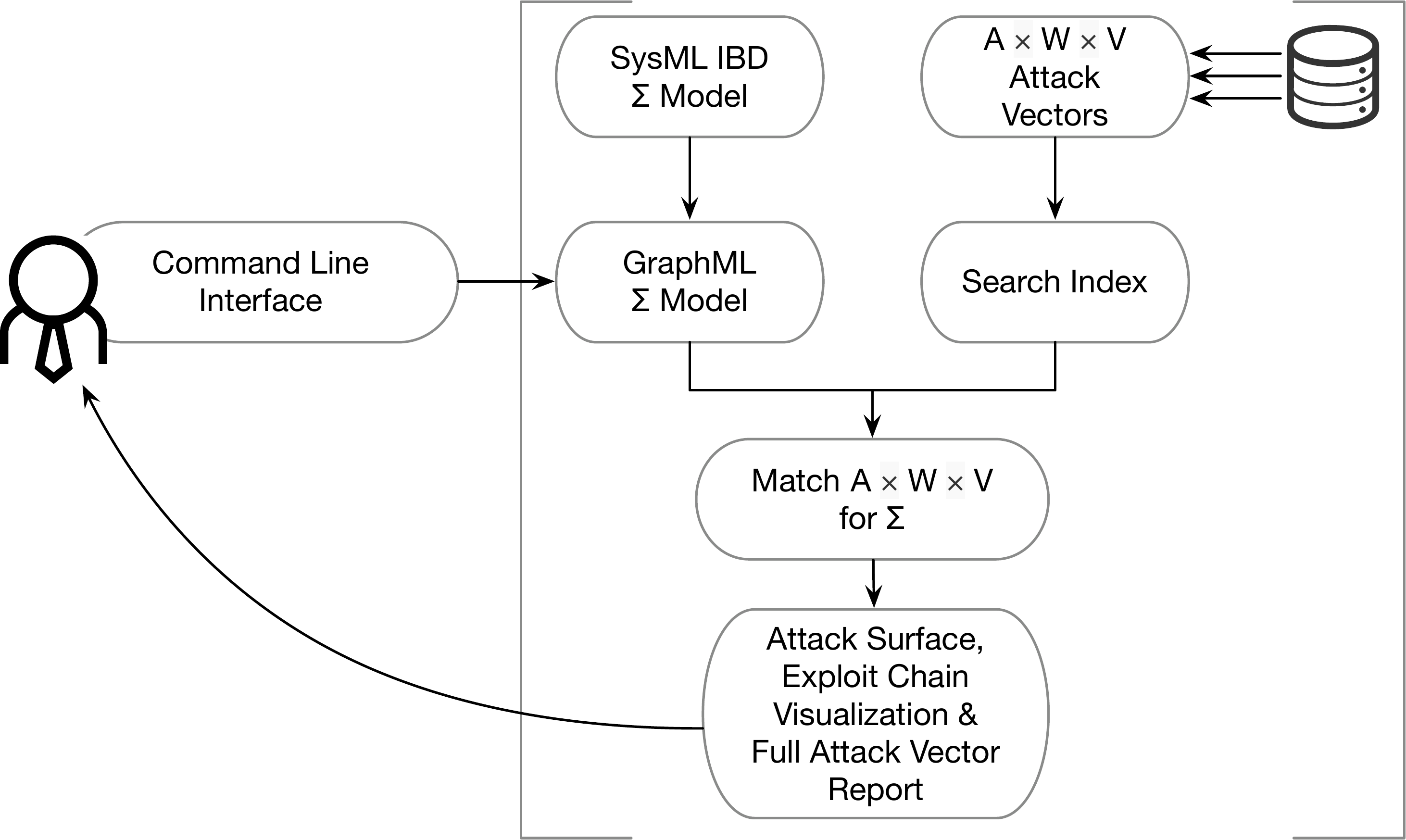}
  \caption{\label{fig:architecture}
    The architecture of CYBOK is modular, meaning that each
    of the functions can be replaced without needing to interfere
    with other functions.
  }
\end{figure}

The overall architecture
of CYBOK is designed
to be modular and robust
with respect to potential architectural changes (Fig.~\ref{fig:architecture}).
For example, the specific implementation
of searching can be changed
without needing to change the rest
of the core tool functionality.

\subsection{Data Extraction, Preprocessing, \& Indexing}

The first step to associating models
to attack vector databases
is to extract
and preprocess the information contained
in several individual databases.
% The first step is
% to download the data
% from MITRE.
An automated mechanism
for downloading the latest set of data
is built into CYBOK because
of the CVE, CWE, and CAPEC update cycle.

Specifically, CAPEC is refactored
and potentially updated every six months
to a year, CWE is extended every one month
to three months,
and CVE adds new entries daily.
By doing automatic updates, the analyst
is sure to have the latest set of data that might inform about new system violations.
% From the lowest levels of abstraction---specific vulnerabilities recorded
% for specific platforms---to the highest more general levels
% of attack patterns and weaknesses.

All database files are encoded in a standard xml file format.
The steps that follow after updating the data files
include preprocessing and constructing the search index
to be used to associate system models to attack vectors.

The preprocessing step extracts the name of the database,
the identification number, name of attack vector, associated attack patterns (if any),
associated weaknesses (if any), associated vulnerabilities (if any),
and the contents; that is the description,
for each entry.
These consist of the full search index schema; that is,
all the information necessary to construct $A \times W \times V$.
% In practice, to achieve this preprocessing step, CYBOK utilizes
% the BeautifulSoup 4.6.0 xml parser.

After preprocessing CYBOK keeps a persistent record
of all attack vectors, $\mathcal{AV}$, by constructing a search index
through the schema defined above.
This allows for efficient information retrieval
of all attack vector information
and avoids having to rebuild $\mathcal{AV}$
for each new search query.
% CYBOK implements search indexing using Whoosh 2.7.4.
% To implement the basic search functionality requires
% using the parsed information to construct a search index.
% The search index is a persistent record
% of all the data
% and is used for efficient information retrieval.
% It also avoids the need
% of parsing the data anew for each search query.
% In CYBOK, search indexing is implemented using Whoosh 2.7.4
% indexing and searching library for Python.

% By extracting, preprocessing,
% and indexing, CYBOK constructs the set
% of attack vectors, \(\mathcal{AV}\),
% to match with system elements.

\subsection{System Models as Graphs}
\label{sec:org748e81a}

Graph structures provide an important view
in a computing system, and
can extend the notion
of violation to more than just a singular view
of components.
% A component individually might be less critical
% for the expected service of the system.
Indeed, the violation of a single component
by an attacker might not be detrimental
to the systems expected service.
However, this component might be connected
to other critical infrastructure.
Therefore, this singular component could be a point of lateral pivot
for an attacker.
This in turn can cause significant malfunction
during operation with catastrophic consequences.
This is how attackers operate
and, therefore, reasoning in graphs of assets provides an attacker's view
to defenders.
For this reason, CYBOK views the system topology; that is, the design artifact, as a graph.
% Practically, this graph structure is encoded in the GraphML file format.

% To import the model, \(\Sigma\), as a graph, CYBOK uses networkx 2.1;
% a Python package for examining graph structures.
% This allows for automatically accessing attributes
% within the model and matching them to the attack vectors
% in \(\mathcal{AV}\).

\subsection{Finding Applicable Attack Vectors from a System Model}

\begin{algorithm}
  \caption{Finding attack vectors}
  \label{lst:attackVectors}
  \begin{algorithmic}[1]
    \Function{Associate}{$\Sigma$, $\mathcal{AV}$}
    \State $\mathcal{R}$ $\gets$ []
    \ForAll {$\textit{desc}(v) \in \mathcal{V}(\Sigma) \wedge \textit{desc}(e) \in \mathcal{E}(\Sigma)$}
    \ForAll {$d \in \mathcal{D}$}
    % get a description d in D from v in V, or e in E
    % using that description you match to terms in a specified schema
    % you return matching instances from db_id.
    \ForAll {$av \in \mathcal{AV}$}
    \If {$d \in av$}
    \State $\mathcal{R}$.append(\{$v \vee e, d, av$\})
    \EndIf
    \EndFor
    \EndFor
    \EndFor
    \State \Return $\mathcal{R}$
    \EndFunction
  \end{algorithmic}
\end{algorithm}

To find applicable attack vectors, CYBOK extracts the descriptive keywords
that define each vertex and edge.
Then, using the descriptive keywords
of the system CYBOK looks at all attack vector entries
from \(\mathcal{AV}\)
to associate the descriptive keywords.
A list of results is returned
for the full system model, \(\Sigma\), including the vertex
or edge the attack vector can exploit, the descriptive keyword, $w_i$, that produced the attack vector,
and the attack vector itself (Algorithm~\ref{lst:attackVectors}).

The search functionality
of CYBOK currently uses a compound word filter.
Other candidates for applying the searching include
n-grams and Shingle filter.

\subsection{Finding Attack Surface Elements}

\begin{algorithm}
  \caption{Finding attack surface elements}
  \label{lst:attackSurface}
  \begin{algorithmic}[1]
    \Function{AttackSurface}{$\Sigma$, $\mathcal{R}$}
    \State $\mathcal{AS}$ $\gets$ []
    \ForAll{$\text{entry\_points}(\textit{desc}(v)) \in \mathcal{V}(\Sigma)$}
    \If  {$\{v, d, \_\} \in$ $\mathcal{R}$}
    \State  $\mathcal{AS}$.append(\{$v$, $d$\})
    \EndIf
    \EndFor
    \State \Return $\mathcal{AS}$
    \EndFunction
  \end{algorithmic}
\end{algorithm}

% CYBOK views the attack surface as any associate specifically
% at the entry point of a vertex description.
CYBOK views the attack surface as any vertex
that has an associated attack vector specifically
at the entry point.
It constructs this set by going through all vertices
and checking if a descriptive keyword, $\text{entry\_points}(w_{i})$,
associatess to an applicable attack vector (Algorithm~\ref{lst:attackSurface}).

\subsection{Finding Exploit Chains}

\begin{algorithm}
  \caption{Finding exploit chains}
  \label{lst:exploitChains}
  \begin{algorithmic}[1]
    \Function{ExploitChains}{$\Sigma$, $\mathcal{R}$, $\mathcal{AS}$, $t$}
    \State $\mathcal{EC}$ $\gets$ $[]$
    % \State admissible\_path $\gets$ $\bot$
    \ForAll {$as \in \mathcal{AS}$}
    \ForAll {$p \in \textit{paths}(as, t, \Sigma)$}
    \ForAll {$v \in p \wedge e \in p$}
    \If  {$\{v \vee e, \_, \_\} \in \mathcal{R}$}
    \State admissible\_path $\gets$ $\top$
    \Else
    \State admissible\_path $\gets$ $\bot$
    \State break
    \EndIf
    \If {admissible\_path $== \top$}
    \State  $\mathcal{EC}$.append(\{$p$\})
    \EndIf
    \EndFor
    \EndFor
    \EndFor
    \State \Return $\mathcal{EC}$
    \EndFunction
  \end{algorithmic}
\end{algorithm}

Exploit chains are paths from a source to a target
that contain violation for every vertex or edge
in that path.
CYBOK finds exploit chains from all elements
of the attack surface, \(\mathcal{AS}\)
to a user input target, \(t\) (Algorithm~\ref{lst:exploitChains}).
These paths are not necessarily the most efficient
or direct paths from the elements
of the attack surface to the given target.
This is because it is often the case
that attackers move laterally from the attack surface
to a specific target without having full observability
of the system.
This way the analyst can be aware
of all paths that are valid based on the system model
and be better informed about potential mitigations.

Furthermore, not all paths from \(\mathcal{AS}\)
to \(t\) are admissible.
Admissible exploit chains require each vertex
and each edge in that path has produced
at least one result from \(\mathcal{AV}\).
Otherwise the path is not fully exploitable based
on evidence and, therefore, does not consist
of an exploit chain under this definition.

\subsection{Visualizations}

\begin{figure}
  \centering
  \includegraphics[width=1\linewidth]{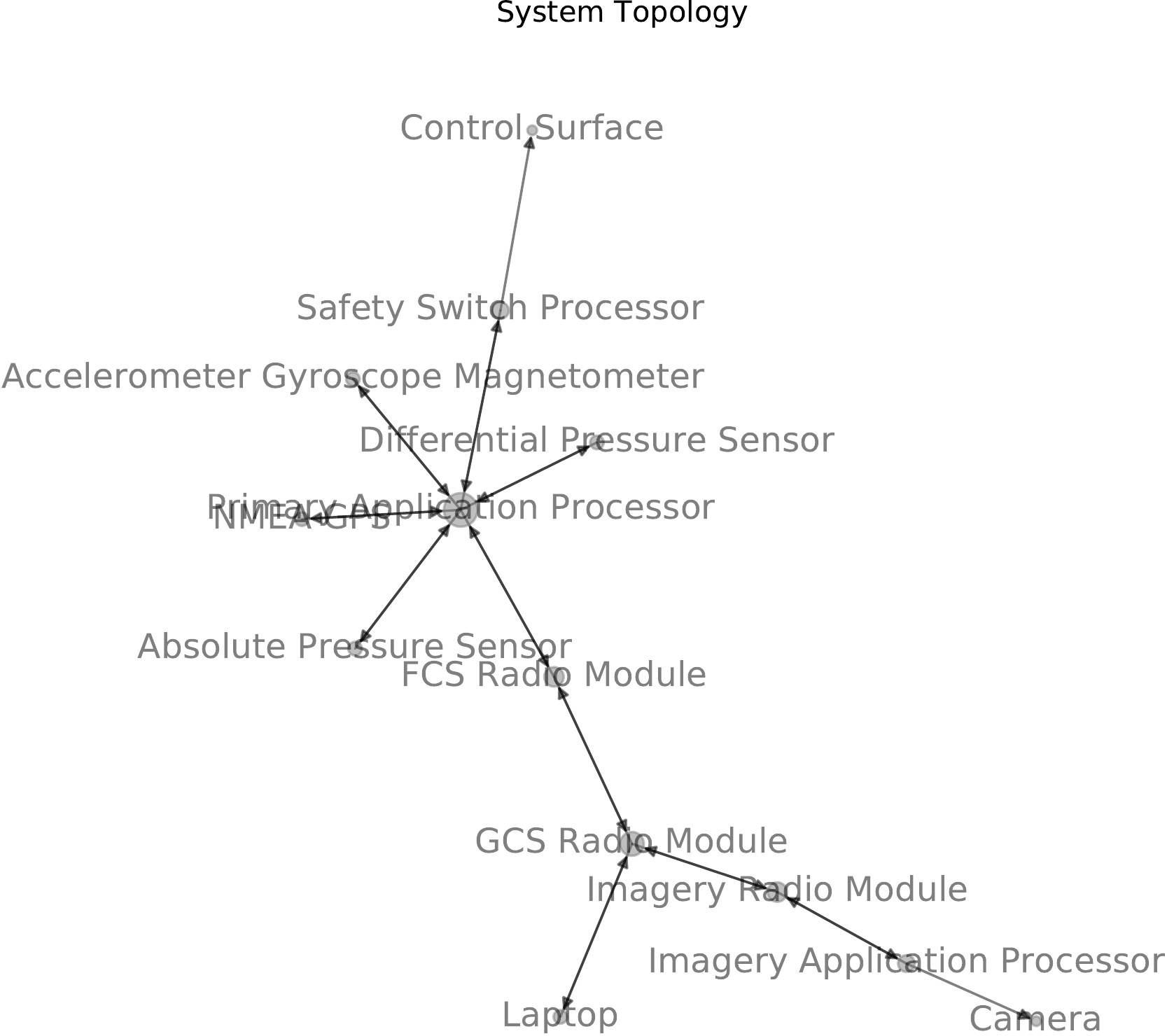}
  \caption{The system topology, \(\Sigma\), shows a static view of the system model.}
  \label{fig:topology}
\end{figure}

To facilitate the analysis of results, CYBOK includes three main visualizations:
(1) the system topology, (2) the system attack surface,
and (3) the system exploit chains.
This is an important feature of CYBOK
because it allows both security analysts
and system designers to project how exploits propagate
over the system model, $\Sigma$.
This way, they can be better informed
about potential mitigation strategies.
For example, changing the definition
of a single element to one that has no recorded attacks
might significantly increase the security posture
of the overall CPS.\footnote{We assume that a component with no recorded attacks is less susceptible to exploitation over one that has a large number of recorded attacks.}

Moreover, a full GUI is developed based on this methodology to implement further interactivity functions
on top of CYBOK~\cite{bakirtzis2018looking}.
This is a natural progression of CYBOK since in-depth analysis requires the analyst
to interact with the data through interactivity functions, for example, filtering,
to facilitate effective exploration
of the diverse types of data input and output to and by CYBOK~\cite{jacobs:2014}.

% CYBOK provides a full textual report
% on the findings based on a model, \(\Sigma\).
% However, the findings can be less than illuminating
% on the actual effects on the system.
% An analyst could spend a lot
% of time examining attack vectors that when viewed holistically
% over the system cause no significant violation of resources.

% For that reason CYBOK provides a set of visualizations
% to assist the analyst in seeing the bigger picture
% of the security posture: topology graph, attack surface graph, exploit chain graph.

% The topology graph can be used
% to verify that the model transformation did not omit important system elements
% or connections of dependence.
% The attack surface graph shows the elements
% and the specific attributes of those elements
% that CYBOK found possible violations
% at the entry point.
% Finally, the exploit chain graph shows all possible admissible exploit chains
% through an animation.

% All these visualizations complement the textual report produced
% by CYBOK.
% In effect, they are used to strengthen the analysis
% and assist the analyst in better deciding mitigation strategies
% in certain areas of the system, \(\Sigma\).

\section{Evaluation}
\label{sec:org111049f}

To evaluate CYBOK we will discuss in some detail
the vulnerability analysis
of one potential design solution for a UAS
that contains a full set of descriptors.
There is ongoing work in applying CYBOK
to several other systems
in the military and nuclear power domain~\cite{beling2019model}.

\subsection{System Model}
\label{sec:orgc3625a5}
% SysML -> graph (what does the graph do for us, that SysML doesn't)
% model of the UAV why is that model a representative model
% give at least an example of attribute...
% have a table of attributes
While modular approaches to flight control systems (FCS)
have been demonstrated to provide flexible choices in hardware~\cite{ward_modular_2014},
it is not currently possible to assess the security
of one design over another before building the system.
By using models of systems it is possible
to assess several system designs
and provide evidence over the use
of one hardware solution over another.
In this work one such hardware solution
is evaluated---through
its system model (Fig.~\ref{fig:topology})---and present the evidence that stems
from assessing the model's security posture using CYBOK.

The potential design solution present in this paper uses several XBee radio modules
to communicate between components,
Dell Latitude E6420 ground control station (GCS) laptop, an ARM STM32F4 primary application processor,
a BeagleBone Black imagery application processor, an ARM STM32F0 safety switch processor, MPU9150 accelerometer, gyroscope, and magnetometer, MS4525DO differential and absolute pressure sensors, a GoPro Hero5 camera,
and an Adafruit Ultimate GPS.
This information is part of the descriptive keywords captured in the vertices
of the model.
Further information is given for both vertices and edges to drive this analysis
per the schema above (Section~\ref{sec:org103a848}).\footnote{The model is publicly distributed~\cite{cybok}.}

\begin{figure}[!t]
  \centering
  \includegraphics[width=1\linewidth]{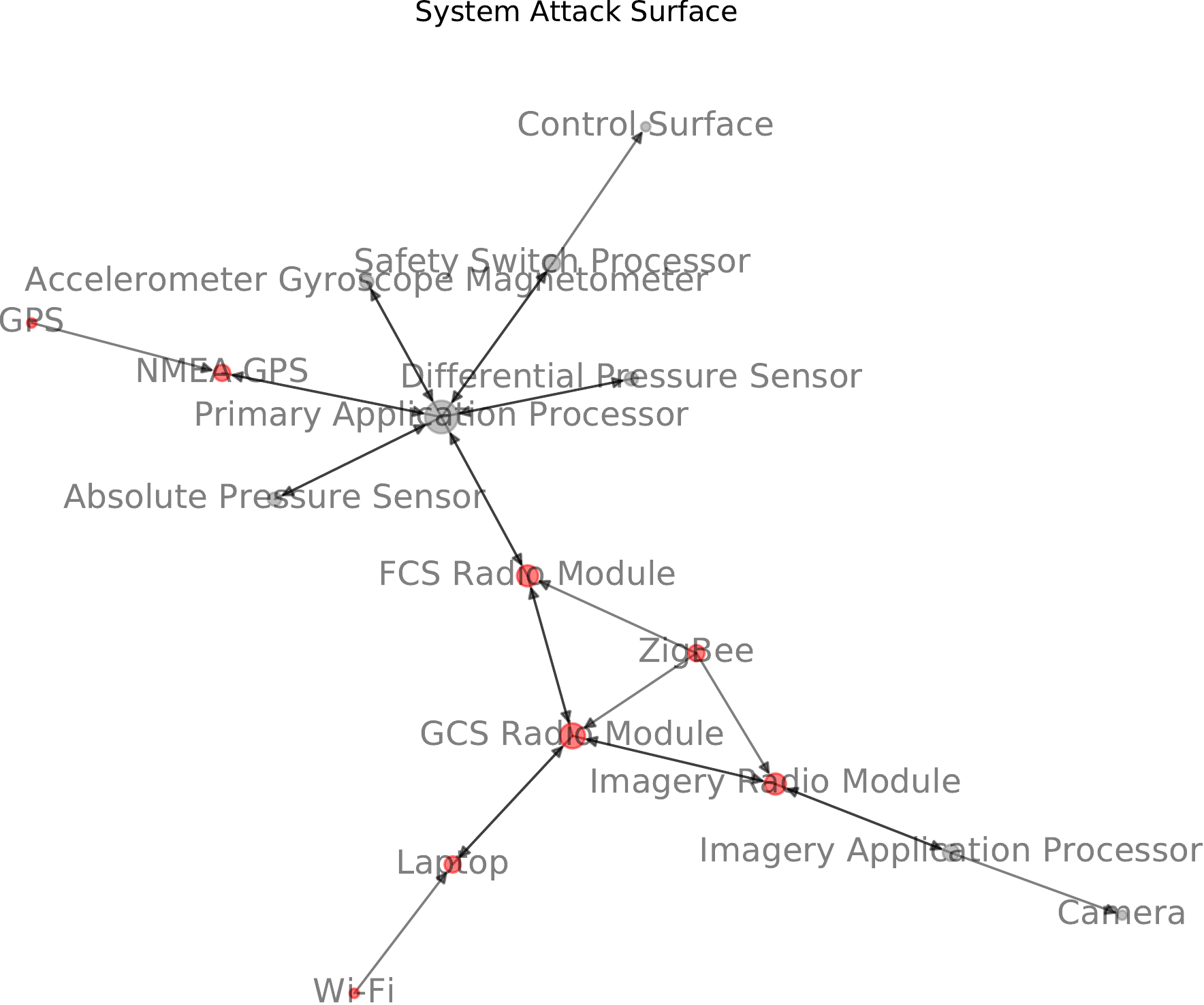}
  \caption{The attack surface, $\mathcal{AS}$, extends the system topology, \(\Sigma\), by adding in the descriptive keywords at the entry point that associate to attack vectors.}
  \label{fig:attack_surface}
\end{figure}

\subsection{Example Analysis}
\label{sec:org7815955}

SysML is used as the modeling language
and tool because it is often familiar to Systems Engineers.
However, CYBOK is not restricted to SysML models, it merely requires a graph representation of the system that includes extra design information (see Bakirtzis et al.~\cite{bakirtzis2018model} and \textsf{graphml\_export}~\cite{bakirtzis:2018b} on how this translation is achieved in practice).
Inputting the UAS system topology to CYBOK first
constructs the attack surface (Fig.~\ref{fig:attack_surface}).
The attack surface is extended to show all the descriptive keywords
at the entry points that attack vectors are found.
For example, by inspection the use of the XBee module
with the ZigBee protocol for all three radio modules can be problematic
because an attacker can exploit the system remotely.
Other such entry points have a different degree of potential exploitation.
It is unlikely that the GPS will be violated but attacks
for GPS exist and, therefore, are reported by CYBOK.
Additionally, the analyst might be aware of hardening techniques
on the Wi-Fi network used by the GCS laptop.
Consequently, the analyst might decide that they consist
of no threat to the systems mission.

From this initial understanding
of the systems security posture
(through its composition and attack surface)
an analyst can further interrogate the model by finding all the potential exploit chains from the attack surface elements
to the primary application processor.
This is because violation of the primary application processor
will cause full degradation
of system functions and, therefore, full mission degradation overall.
Specifically, an analyst might want to examine a potential exploit chain
stemming from the XBee element of the attack surface.
By providing a target, \(t\), CYBOK finds the admissible paths
and, therefore, exploit chains
from the imagery radio module to the primary application processor.
This path is admissible if each vertex \emph{and} each edge within
that path has produced evidence; that is, attack vectors.

% As of now a specific set
% of attack vectors has not been considered.
% % Instead, we examine the projection
% % of the results over the system topology.
% To better understand the feasibility and existence
% of those attack; that is, to find evidence, there are two options:
% (1) to sort through the textual report
% produced by CYBOK on the command line
% or (2) use the GUI dashboard to clearly see the relationships between attack vectors and how those attack vectors associate to the system topology.
% The results here are illustrated through the security analyst dashboard.

\begin{figure}[!t]
  \centering
  \includegraphics[width=1\linewidth]{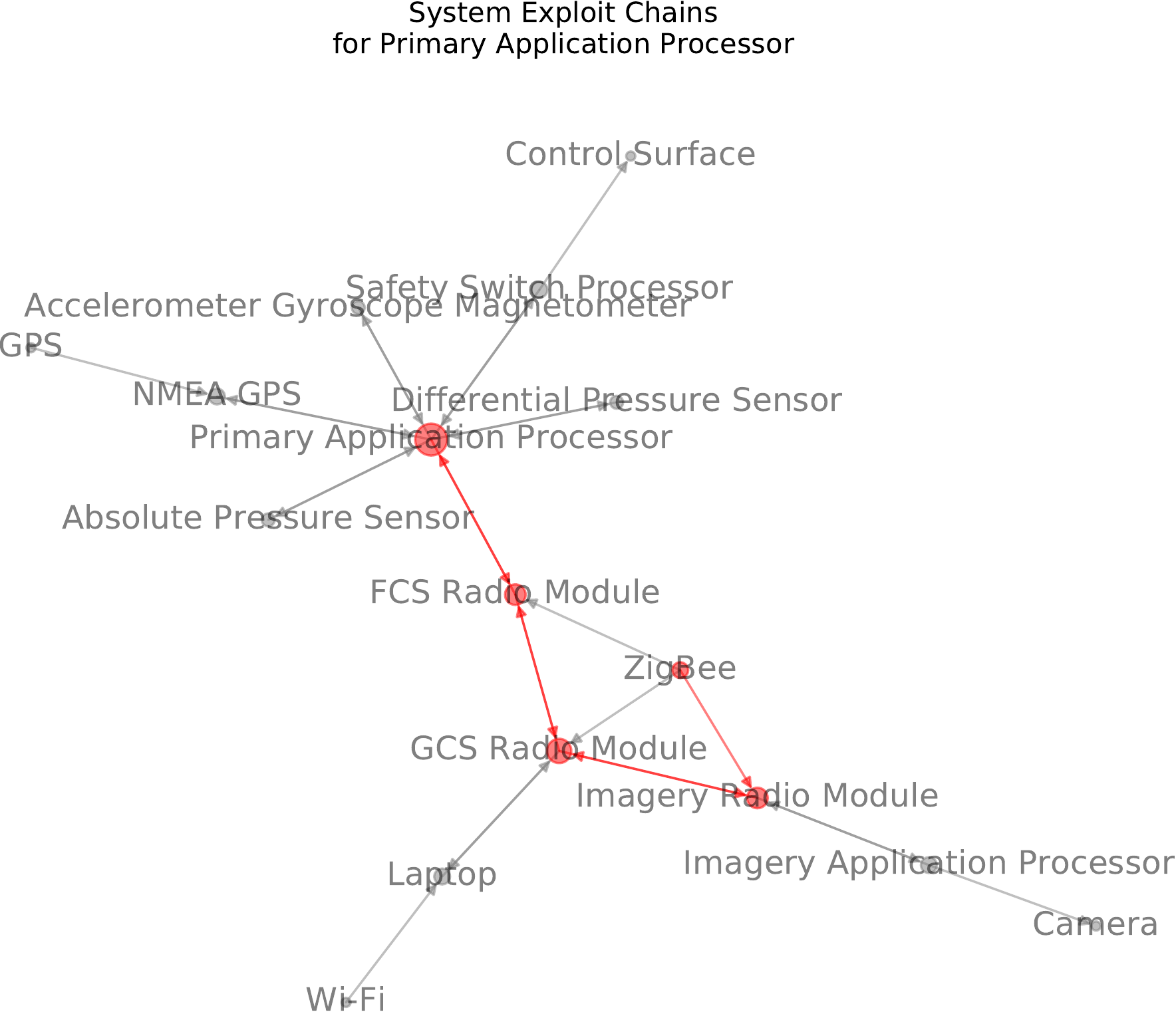}
  \caption{Exploit chains, $\mathcal{EC}$, show a possible lateral paths an attacker might take over the system topology to reach a specific system element. This is but one example of such exploit chain from the attack surface, \(\mathcal{AS}\), to some target \(t\)---in this case the primary application processor.}
  \label{fig:exploit_chain}
\end{figure}

% The dashboard consists of two main panes.
% One showing the system topology, attack surface, and exploit chains
% and one showing the associated attack vector space.
% A third pane is used to record specific attack vectors
% that an analyst might want to project over the system topology
% or further examine and report to other stakeholders
% to discuss mitigative actions.
Examining the results produced by CYBOK we find the following three associated entries: \texttt{CAPEC-67} ``String Format Overflow in syslog(),''
\texttt{CWE-20} ``Improper Input Validation,''
and \texttt{CVE-2015-8732} a specific attack
on the ZigBee protocol used by XBee that allows remote attackers
to cause a denial of service (DOS) via a crafted packet.
Further, for the edges from the radio module to the primary application processor
there are the following attack vectors produced by CYBOK, \texttt{CVE-2013-7266},
which is a specific attack that takes advantage
of not ensuring length values matching the size of the data structure, \texttt{CWE-20} ``Improper Input Validation,'' \texttt{CWE-789} ``Uncontrolled Memory Allocation,'' \texttt{CWE-770} ``Allocation of Resources without Limits or Throttling,'' and \texttt{CAPEC-130} ``Excessive Allocation.''
Finally, the primary application processor uses the I2C and RS-232 protocols
to communicate with the rest of the hardware (these are descriptive keywords contained
in the edges of the graph), which produce the following, \texttt{CAPEC-272} ``Protocol Manipulation''
and \texttt{CAPEC-220} ``Client-Server Protocol Manipulation.''
All this information is used as evidence for the feasibility of one exploit chain
from the attack surface to the primary application processor (Fig.~\ref{fig:exploit_chain}).

By projecting the attack over the system structure it is evident when the same attacks are applicable to several parts of the system.% (Fig.~\ref{fig:dashboard}).
This is important because attackers contain a specific skill set
and they do not usually deviate from it if not necessary.

\begin{table*}[!t]
  \caption{A fragment of relevant results for the UAS as produced by CYBOK.}
  \label{tab:results}
  \centering
  \begin{tabular}{@{}lll@{}}
    \toprule
    Model Element                & Attack Vector & Description                                                                 \\ \toprule
    Radio Modules                 & CVE-2015-6244 & Relies on length fields in packet data, allows attacks from crafted packets \\
                                 & CWE-20        & Improper input validation                                                   \\
                                 & CAPEC-67      & String format overflow in syslog()                                          \\ \midrule
    NMEA GPS                      & CAPEC-627     & Counterfeit GPS signals                                                     \\
                                 & CAPEC-628     & Carry-Off GPS attack                                                        \\ \midrule
    Primary Application Processor & CVE-2013-7266 & Does not ensure length values match size of data structure                  \\
                                 & CWE-20        & Improper input validation                                                   \\
                                 & CWE-789       & Uncontrolled memory allocation                                              \\
                                 & CWE-770       & Allocation of resources without limits or throttling                        \\
                                 & CAPEC-130     & Excessive allocation                                                        \\ \midrule
    I2C \& RS-232 Protocols       & CAPEC-272     & Protocol manipulation                                                       \\
                                 & CAPEC-220     & Client-server protocol manipulation                                         \\ \midrule
    Imagery Application Processor & CWE-805       & Buffer access with incorrect length value                                   \\
                                 & CAPEC-100     & Overflow buffers                                                            \\ \midrule
    Safety Switch Processor       & CWE-1037      & Processor optimization removal or modification of security-critical code    \\ \midrule
    Laptop                        & CAPEC-615     & Evil twin Wi-Fi attack                                                      \\
                                 & CAPEC-604     & Wi-Fi jamming                                                               \\ \midrule
    Camera                        & CVE-2014-6434 & Allows remote attackers to execute commands in a restart action \\ \bottomrule
  \end{tabular}
\end{table*}

% \begin{figure*}
%   \centering
%   \includegraphics[width=1\linewidth]{./figures/dashboard}
%   \caption{The dashboard allows examining the system and its associated attack vector space concurrently. It also lets analysts project attack vectors over the system structure, filter the attack vector space based on component or keywords, record attack vectors, and also change the descriptive keywords within the system model to apply a ``what if'' analysis.}
%   \label{fig:dashboard}
% \end{figure*}

Additionally, CYBOK allows flexible ``what-if'' analysis
by changing the descriptive keywords in the model.
For example, by changing the radio module definition from XBee
using the ZigBee protocol to some other radio module offered in the market
might exclude it from the attack surface.
Since a larger attack surface implies more access points
and, therefore, a less secure system an analyst might decide
to propose changing the design of the system.

A full analysis consists of first identifying the important elements
of the system; that is, the assets that might require protections.
This might be informed from the outputs produced by CYBOK
or from expert input and information elicitation.
Then, of filtering the large space
of attack vectors that associate to the model
to find the most relevant and strong evidence (Table~\ref{tab:results}).
This evidence is what ultimately informs other stakeholders,
such that they can devise mitigative actions---changing the system solution
to conform to mission needs, erecting security barriers at strategic points,
or applying resilience solutions during operation.

\section{Related Work}

Little research has been done
for evidence-based security assessment
in a model-based setting.
Usually work in this area requires \textit{transcribing} already known vulnerabilities
to a modeling tool and assessing if it might apply to a system design.
Instead, the aim of this work is to employ models
that can---by their fidelity---immediately produce a large number
of potential attack vectors.
These attack vectors stem from the model itself
and are not informed from some a priori security knowledge.
% As an example of the former approach, Ouchani et al. \cite{Ouchani2014pcs} propose a framework
% which can automatically apply attacks to SysML models.
% However, it requires not only knowing the specific
% attacks one has to check based on the given system,
% but also requires manually transcribing the attack in SysML.
% In comparison, CYBOK within a larger modeling framework
% only requires a certain set of characteristic desciptors
% to produce a vulnerability report.
% This report might be significantly larger
% (from matching the model descriptor to CVE),
% CYBOK can report abstraction of classes of attacks---through
% the attack patterns and weaknesses associated with them.
% This sets a practical use case for CYBOK.

We acknowledge
that some current attack vector search tools could be repurposed
for model-based systems engineering.
One such search tool is cve-search~\cite{cve-search}.
However, cve-search cannot input a system model.
It only provisions security datasets
in one search engine.
It is also limited with respect
to visualization techniques.

Noel et al.~\cite{noel2016cygraph} propose CyGraph which also is based
on a graph-based understanding of the system
but this work fundamentally differs in scope (mainly targets traditional networked systems)
and approach (uses a traditional notion of attack graphs).

Adams et al.~\cite{adams2018selecting} propose topic modeling
for finding applicable attack vectors given a system model.
However, they only examine CAPEC as a potential source
of attack vectors, which is necessary but insufficient.

Ford et al.~\cite{ford2013implementing} propose using the ADVISE security methodology~\cite{lemay2011model}
on top of the M{\"o}bius tool~\cite{courtney2006data}
to provide an attackers's view.
However, the quantitative analysis is based
on profiling and modeling attacker actions.
The framework is largely unaware of a specific system model
that could be used to implement a realized system.

The analysis presented in this paper is qualitative.
This is because quantitative information for cyber-physical attacks is limited
and ultimately expert input is necessary to understand what it means
for a metric to show that a system is more susceptible to attacks
over another.
For example, a number of quantitative approaches incorporate CVSS
as a potential metric for risk~\cite{frigault2008measuring,houmb2010quantifying,wang2015data}.
But, CVSS only defines severity of a given vulnerability and not risk~\cite{CVSS,collier2014cybersecurity}.

In general, to the best of the authors' knowledge, there is no direct comparison between the work in this paper and existing work in the literature. It is challenging to do a direct comparison with any existing models because previous work is based on an already implemented system or does not apply attack vector information directly to the model.

% \section{Context}

% The limitations are:
% more modeling effort
% requiring an initial design through some expertise
% a specific model (graph) is required.

% Nevertheless, to out knowledge this is the first work
% that provides \emph{realistic} analysis at the design phase.

\section{Conclusion}
\label{sec:org03d22dc}

In this paper we propose a method
and implement a tool to support this method, CYBOK,
that is able to find associated attack vectors
given a sufficient system model.
CYBOK provides flexibility in modifying the system model
to represent different design solutions
that implement the same desired behaviors.
Therefore, moving security analysis earlier in the systems lifecycle---particularly
at the design phase---and, therefore, building systems with security by design.
Two important metrics are used
for assessing a systems security posture; the attack surface and exploit chains.
The results of this method and toolkit is illustrated and evaluated
using a UAS---an important area
for secure system design because exploits can cause hazardous behavior.

As a final observation we note the experience
of using a systematic, model-driven process
to conduct attack vector analysis often yields more information
than just quantifying the vulnerability aspects of the system.
The process itself is an iterative learning experience, allowing circumspection
into how a system behaves in response to potential exploits.
% To assist with this issue we present a GUI for CYBOK
% that provides useful interactivity functions
% to security analysts and system designers.

\section{Acknowledgments}

This material is based upon work supported in part by the Center for Complex Systems and Enterprises at the Stevens Institute of Technology and in part by the United States Department of Defense through the Systems Engineering Research Center (SERC) under Contract HQ0034-13-D-0004. SERC is a federally funded University Affiliated Research Center managed by Stevens Institute of Technology. Any opinions, findings and conclusions or recommendations expressed in this material are those of the authors and do not necessarily reflect the views of the United States Department of Defense.

% \balance
\bibliographystyle{IEEEtran}
\bibliography{manuscript}

\end{document}